\documentclass[aps,prl,preprint,superscriptaddress,showpacs]{revtex4}

\usepackage{graphicx}
\usepackage{amsmath}
\usepackage{graphicx}
\newcommand{\figwidth}{0.47\textwidth}

\begin{document}

\title{Magnetic vortex as a ground state  for sub-micron
antiferromagnetic particles }

\author{A.~Yu. Galkin}
\affiliation{Institute of Magnetism, 03142 Kiev, Ukraine}
\affiliation{Institute of Metal Physics, 03142 Kiev, Ukraine }

\author{B.~A. Ivanov}
\email{bivanov@i.com.ua} \affiliation{Institute of Magnetism, 03142
Kiev, Ukraine}
\affiliation{National Taras Shevchenko University of
Kiev, 03127 Kiev, Ukraine}

\date{\today}

\begin{abstract}
For submicron particles shaped as any axisymmetric body and made
with standard canted antiferromagnet like hematite or iron borate,
the ground state may comprise of a magnetic vortex with
topologically non-trivial distribution of the sublattice
magnetization $\vec{l}$ and planar coreless vortex-like structure
for the net magnetization $\vec{M}$. For antiferromagnetic particles
in the vortex state, in addition to low-frequency modes, there are
high frequency modes with frequencies over the range of hundreds
gigahertz, including a mode localized in the region of the radius
30-40 nm near a vortex core.
\end{abstract}

\maketitle

The magnetic properties of submicron ferromagnetic (FM) particles of
the shape of circular cylinder (magnetic dots) are today under great
consideration mainly due to their potential applications
\cite{Skomski03}. Circular dots possess the equilibrium magnetic
configuration which corresponds to a vortex structure just above the
single domain length scale, with the radius $R > R_{\mathrm{crit}}
\sim$ 100-200 nm. The FM vortex state consists of an in-plane
flux-closure magnetization distribution and a central core with
radius of order of 20-30 nm magnetized perpendicularly to the dot
plane. Reducing of the magnetostatic energy comes at the cost of a
large exchange energy near the vortex core, as well as the
magnetostatic energy caused by the core magnetization. For data
storage purposes, vortex state is of considerable interest, because
the low magnetic stray field reduce the particles interaction and
leads to a high magnetic stability of the data written. The magnon
modes for dots in virtue of effects of spatial quantization possess
a discrete spectrum. The possibility to manage localization and
interference of magnons spawned an idea of so-called
\emph{magnonics} based on usage of these modes for developing a new
generation of microwave devices with submicron active elements
\cite{magnonics}. These particles also provide an ideal experimental
system for studying static and especially dynamic properties of
relatively simple topologically non-trivial magnetic structures
which are fundamentally interesting objects in the research area of
magnetism.

All previous studies of magnetic vortices caused by magnetic dipole
interaction were carried out on the magnetic particles made with
soft FM with high magnetization $M_s $ like permalloy with $4\pi M_s
\sim 1$ T. In this Letter we have shown that the vortices can be a
ground state for sub-micron particles made with another important
class of magnetic materials, antiferromagnets (AFM), with easy-plane
(EP) anisotropy and Dzyaloshinskii-Morya interaction (DMI).

For AFM, exchange interaction between neighboring spins facilitates
antiparallel spin orientation, which leads to the structure with two
antiparallel magnetic sub-lattices, $\vec {M}_1 $ and $\vec {M}_2 $,
$\vert \vec {M}_1 \vert =\vert \vec {M}_2 \vert =M_0 $. As typical
AFM we can mention hematite $\alpha$-$\mbox{Fe}_2 \mbox{O}_3 $, iron
borate $\mathrm{FeBO}_3 $, and orthoferrites, see
\cite{Bar-springer}. These materials are characterized by high
temperatures of magnetic ordering and have unique physical
properties: orthoferrites and iron borate are transparent in optical
range and have a strong Faraday effect, the magnetoelastic coupling
is quite high in hematite and iron borate \cite{Bar-springer}. They
possess small but non--zero net magnetization caused by a weak
non-collinearity of sublattices (sublattice canting) originated from
DMI.

To describe the structure of AFM, it is convenient to introduce
irreducible combinations of the vectors $\vec {M}_1 $ and $\vec
{M}_2 $, the net  magnetization $\vec {M}=\vec {M}_1 +\vec {M}_2
=2M_0 \vec {m}$ and the vector of sublattice magnetization, $\vec
{l}=(\vec {M}_1 -\vec {M}_2 )/2M_0 $. The vectors $\vec {m}$ and
$\vec {l}$ are subject to constraint $(\vec {m} \cdot \vec {l})=0,
\quad \vec {m}^2+\vec {l}^2=1$.  As $\vert \vec {m}\vert \ll 1$, the
vector $\vec {l}$ could be considered as a unit vector. The mutual
orientation of sublattices is determined by a sum of the energy of
uniform exchange $W_\mathrm{ex}=H_\mathrm{ex} M_0 \vec {m}^2$, and
the DMI energy, $W_{\mathrm{DM}}=2M_0 H_\mathrm{D} (\vec {d} \cdot
(\vec {m}\times \vec {l}))$, the unit vector $\vec {d}$  is directed
along the symmetry axis of a magnet. The parameters $H_{\mathrm{ex}}
\sim 3\cdot 10^2-10^3 $ T and $H_\mathrm{D}  \sim 10$ T are exchange
field and DMI field, respectively. Using this energy and dynamical
equations for $\vec {M}_1 $ and $\vec {M}_2 $, one can find,
\begin{equation}
\label{eq5} \vec {M}=M_{\mathrm{DM}} \left(\vec {d}\times \vec
{l}\right)+ \frac{2M_0}{\gamma H_{\mathrm{ex}}}\left(\vec {l} \times
\frac{\partial \vec {l}}{\partial t}\right) ,\;M_{\mathrm{DM}} =
\frac{2H_{\mathrm{D}} M_0 }{H_{\mathrm{ex}}} \, ,
\end{equation}
where the first term gives the static value of AFM net magnetization
$M_{\mathrm{DM}} $, comprising a small parameter, $H_\mathrm{D}
/H_{\mathrm{ex}} \sim 10^{-2}$, second term describes the dynamic
canting of sublattices, see for details \cite{Bar-springer}.
$M_{\mathrm{DM}} $ is much smaller than $M_0 $ or the value of $M_s$
for typical FM, but the role of the magnetostatic energy caused by
$M_{\mathrm{DM}} $  could be essential, and could lead to the
appearance of a domain structure for AFMs \cite{book,Bar-springer}.
We will show that for formation of equilibrium vortices in AFM even
have some advantage compared with soft FM.

The dynamical properties of AFM are essentially different comparing
with FM. A spin dynamics of an AFM can be easily described in the
framework of so-called sigma-model equation ($\sigma$-ME), a
dynamical equation for the vector ${\rm {\bf l}}$ only, with the
magnetization $\vec {M}$ being a slave variable \cite{Bar-springer}.
In contrast to the Landau -- Lifshitz equation for a FM
magnetization, the $\sigma$-ME contains a dynamical term with a
second time derivative of $\vec{l}$, combined with gradients of
$\vec{l}$ in the Lorentz-invariant form $d^2\vec l / dt^2 -
c^2\nabla ^2 \vec {l}$. For this reason, for AFM two magnon branches
(instead of one, for FM) exist. The chosen speed $c = \gamma
\sqrt{AH_\mathrm{ex}/M_0} $ plays roles of both magnon speed and the
limit speed of domain walls, it is determined by exchange
interaction only and attains tens km/s, $c \simeq 1.4 \cdot 10^4$m/s
for iron borate and $c \simeq 2 \cdot 10^4$ m/s for orthoferrites
\cite{Bar-springer}. For both modes the elliptic polarization of the
oscillations of $\vec M_1$ and $\vec M_2$ is such that  the
oscillations of the vector $\vec l$ have linear polarization
\cite{Bar-springer}. For EP AFM these two branches are low-frequency
quasi-ferromagnetic (QFM) branch and a high-frequency
quasi-antiferromagnetic (QAF) branch. QFM magnons involves the
oscillations of the vectors $\vec l$ and $\vec M$ in the EP, with
weak deviation of $\vec M $ from the EP caused by last summand in
\eqref{eq5}.  The second QAF branch corresponds to the out-of-plane
oscillations of $\vec l$ with the dispersion law $\omega
_{\mathrm{QAF}} (\vec {k}) = \sqrt {\omega _\mathrm{g}^2 + c^2\vec
{k}^2} $, where $\vec {k}$ is the magnon wave vector. The gap of QAF
branch, $\omega _\mathrm{g} = \gamma \sqrt {2H_{\mathrm{ex}}
H_\mathrm{a} } $ contains large value $H_{\mathrm{ex}}$ and attains
hundreds GHz. Thus both magnon frequency and domain wall speed for
AFM dynamics, comparing with FM, contain a large parameter
$\sqrt{H_\mathrm{ex}/H_\mathrm{a}}\sim 30-100$, $H_\mathrm{ex}$  and
$H_\mathrm{a}$ are exchange field and anisotropy field,
respectively, which can be referred as \emph{exchange amplification}
of dynamical parameters of AFM. The frequencies of AFM magnon modes
$\omega_\mathrm{g}$ reaches hundreds GHz, with values 170 GHz for
hematite, 100-500 GHz for different orthoferrites and 310 GHz for
iron borate \cite{data}. Recent studies showed a possibility to
excite spin oscillations of non--small amplitude for orthoferrites
\cite{Kimel} and iron borate \cite{Kalash} with the use of
ultra-short laser pulses.

Spin distribution for AFM can be described by the energy functional
of the form $W[\vec {l}] + W_m $. Here $W[\vec {l}] $ describes the
energy of non-uniform exchange and the anisotropy energy through
only the vector $\vec {l}$; $W_m $ is magnetic dipole energy,
\begin{equation}
\label{eq3} W_m =-\frac{1}{2} \int {\vec {M}\vec {H}_m } d^3x,
\end{equation}
where $\vec {H}_m $ is demagnetization field caused by AFM
magnetization \eqref{eq5}. The sources of $\vec {H}_m $ can be
considered as formal ``magnetic charges'', both volume charges equal
to $\mbox{div} \vec {M}$ and surface charges equal $-\vec {M}\cdot
\vec {n}$, see monographs \cite{SW,book} for general consideration
and \cite{IvZasAll} for application to vortices.

For a pure uniaxial model of an AFM which is applicable for hematite
and iron borate, and to some extend for orthopherites, $W[\vec {l}]
$ can be presented as,
\begin{equation} \label{eq6}
W[\vec {l}]=\frac{1}{2}\int {[A(\nabla \vec {l})^2+K\cdot l_z^2 ]}
d^3x \, ,
\end{equation}
where $A$ is the non-uniform exchange  constant and $K$ is
anisotropy constant, the  $xy-$plane is the EP for spins. Variation
of the energy $W[\vec {l}]$ gives a general  two-dimensional (2D)
vortex solution for the vector $\vec {l}$ of the form
\begin{equation}
\label{eq1} \vec {l}=\vec {e}_z \cos \theta +\sin \theta [\vec {e}_x
\cos (\chi +\varphi _0 )+\vec {e}_y \sin (\chi +\varphi _0 )]
\end{equation}
where  $\theta=\theta (r)$, $r$ and $\chi $ are polar coordinates in
an EP of a magnet, the vector $\vec {e}_z $ is the hard axis, the
value of $\varphi _0 $ is arbitrary, see \cite{AFMvort91,ISchMW}.
The function $\theta (r)$ exponentially tends to $\pi $/2 at $r \gg
l_0 $, with the characteristic size $l_0 =\sqrt {A/K} $, and in the
center of the vortex (at $r$ = 0) $\sin \theta (0)=0$. Near the
vortex core $\vec l$ deviates from the EP which leads to the loss of
the anisotropy energy. The state (\ref{eq1}) is non-uniform, what
corresponds to the loss in the exchange energy. Therefore, for the
EP model without taken into account the magnetic dipole interaction
the appearance of a vortex costs some energy, i.e. the vortex
corresponds to excited states of AFM. Vortex excitations are
important for description of thermodynamics of 2D AFM, see
\cite{AFMvort91}.

For small particles made with canted AFM the energy loss caused by a
vortex can be compensated by the energy of magnetic dipole
interaction. To explain this, note that for a uniform distribution
state the contribution $W_m $ unavoidably results in a loss of the
system energy, which is proportional to the particle volume $V$
\cite{SW,book}. The energy of the uniform state could be estimated
as
\begin{equation}\label{Ehomog}
E^{(\mathrm{homog})}=2\pi NM_{DM}^2 V=2\pi NM_0^2 (2H_\mathrm{D}
/H_e )^2V,
\end{equation}
where $N$ is the effective demagnetizing factor in the direction
perpendicular to the particle axis \cite{SW,book}. In contrast, for
the vortex state \eqref{eq1} with a chosen value of $\sin \varphi _0
=0$, with $\vec M \propto (\vec {d} \times \vec {l})$, one can find
\begin{equation} \label{eq7}
\vec {M}=\sigma  M_\mathrm{DM}\cdot \sin \theta (-\vec {e}_x \sin
\chi +\vec {e}_y \cos \chi ),
\end{equation}
where $\sigma =\cos \varphi _0 =\pm 1$. A unique property of the
state \eqref{eq7} is that it can also \emph{exactly} minimize the
energy of the magnetic dipole interaction $W_m$, giving $\vec {H}_m
=0$ in the overall space. Indeed, the projection of $\vec {M}$ on
the lateral surface of any axisymmetric body with the symmetry axis
parallel to $z$-axis, as well as $\mathrm{div} \vec {M}$, equal to
zero. Moreover, in virtue of symmetry of DMI $(\vec {M} \cdot \vec
{d})=0$, the distribution of the magnetization $\vec {M}$
(\ref{eq7}) is \emph{purely planar} (in contrast to $\vec {l}$) and
the out of plane component of $\vec {M}$ is absent. In the vicinity
of the vortex core the length of vector $\vec {M}$ decreases,
turning to zero in the vortex center, see. Fig.~1. Such feature is
well known for domain walls in some orthoferrites \cite{BulGins}.
Thus the AFM vortex is the unique spin configurations which do not
create demagnetization field in a singly connected body (for FM with
$|\vec{M}|=\mathrm{const}$ a configuration with $\vec H_m \equiv 0$
is possible only for a magnetic rings having the topology of torus).

\begin{figure}[htbp]
\includegraphics[width=\figwidth]{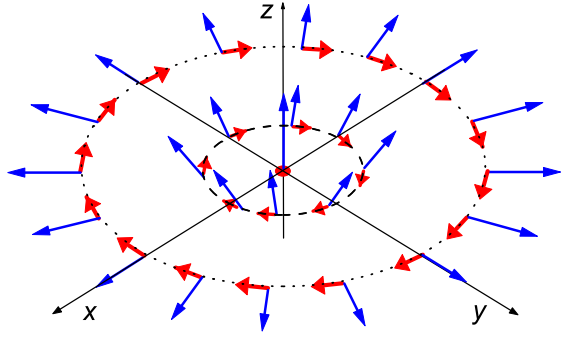}
\caption{\label{fig1} (Colour online) The structure of the AFM
vortex (schematically). The vectors $\vec {l}$ (thin blue arrows)
and $\vec {M}$ (thick red arrows; present not in scale) are depicted
for the area of the core (dashed circle) and far from it (dotted
circle); the full red dot in the origin indicates the value $\vec
{M}=0$ for the state with $\vec {l}$ perpendicular to the plane.}
\end{figure}

Let us compare the energies of the vortex state and the uniform
state for the AFM particle  shaped as a cylinder with the height $L$
and the radius $R$. For the vortex state $\vec {H}_m =0$ overall the
volume of the particle, and the vortex energy is determined by the
simple formula \cite{Bar-springer}
\begin{equation}\label{Evort}
E_{\mathrm{v}}=\pi AL\ln \left( {\rho \frac{R}{l_0 }} \right),
\end{equation}
where $\rho \approx 4.1$ is the numerical parameter. For long
cylinder with $L \gg R$ the value of  $N\simeq 1/2$ and the vortex
state becomes favorable if the radius $R$ exceeds some critical
value $R_\mathrm{crit} $,
\begin{equation}
\label{eq9} R\ge R_{\mathrm{crit}} = 2 l_{\mathrm{dip}} \sqrt {\ln
\left(  \frac{l_{\mathrm{dip}}}{l_0} \right)}, \
l_{\mathrm{dip}}=\sqrt {\frac{A}{4\pi M_{\mathrm{DM}}^2 }},
\end{equation}
where $l_{\mathrm{dip}}$ determines the spatial scale corresponding
to the magnetic dipole interaction. Note that this quantity
comprises a large parameter $H_e /H_\mathrm{D} \sim 30 - 100$, and
$l_{\mathrm{dip}} \gg l_0$. In the case of a thin disk, $ L \ll R$
the demagnetization field energy could be revealed as
$E^{(\mathrm{homog})}=2\pi RL^2M_{DM}^2 \ln (4R/L)$ \cite{IvZasAll},
and the vortex state is energetically favorable for $RL \ge
(RL)_{\mathrm{crit}} =2 l_{\mathrm{dip}}^2 $.

For concrete estimations we take the parameters of iron borate,
$A=$0.7$\cdot 10^{-6}$ erg/cm, $K=$4.9$\cdot 10^{6}$ erg/cm$^3$ and
$4\pi M_{\mathrm{DM}} =$120 Oe. Then we obtain that $l_0 =3.8$ nm,
i.e. the core size is of the same order of magnitude as for typical
FM (for permalloy $l_0 =4.8$ nm). The value $l_{\mathrm{dip}} $ is
essentially higher, for iron borate $l_{\mathrm{dip}} =220$ nm.
Combining these data one finds for the long cylinder
$R_{\mathrm{crit}} =$0.9 $\mu$m. For a thin disk sample the
characteristic scale has submicron value, $\sqrt
{(RL)_{\mathrm{crit}} } =0.4$ $\mu$m. The similar estimations are
obtained for orthoferrites, and somewhat higher values for hematite.
Thus, despite the fact that characteristic values for the dipole
length $l_{\mathrm{dip}} $ for FM and AFM  differ hundredfold, the
characteristic critical sizes differ not so drastically (for
permalloy $R_\mathrm{crit} \sim$ 100-200 nm). It is caused by the
aforementioned fact that  the magnetic field created by the vortex
core is completely absent for the AFM vortex. The situation here is
common to that for FM nanorings, where the vortex core is absent and
the vortex state is more stable that for FM dots.

Despite that the vortex core size in FM dots is rather small, the
core contribution to $W_m$ for ferromagnetic particles of rather big
radius $R\ge 0.5\;\mu $m is negligible, but it becomes essential for
small particles with $R$ close to the critical size. Note as well
that the vortex core magnetic field in the FM destroys a purely 2D
distribution of $\vec {M}$ like (\ref{eq1}), and the core size
changes over the thickness of the particle. For the AFM vortex the
value of $\vec {H}_m $ equals exactly zero, and truly 2D
distribution of $\vec {l}$ and $\vec {M}$, independent on a
coordinate $z$ along the body axis, see (\ref{eq7}, \ref{eq1}), is
possible.

Since magnon spectra of bulk FM and AFM differ significantly, one
can expect an essential difference for magnon modes for vortex state
AFM  and FM particles. Remind briefly the properties of normal modes
for disk shaped vortex state FM particles. For such samples, the
presence of discrete spectrum of magnon modes, characterized by the
principal number (the number of nodes) $n$ and the azimuthal number
$m$, is well established \cite{GiovanniniExper,BuessExper}. This
spectrum includes a single low-frequency mode of precessional motion
of a vortex core ($n = 0$, $m = 1)$ which has the frequency in
subGHz region \cite{GusIv}, a set of radially symmetrical modes with
$m = 0$ \cite{mZero}, and also a system of slightly splinted
doublets with the azimuthal numbers $m = \pm \vert m\vert $, with
frequencies $\omega _{|m|, n} \neq \omega _{-|m|, n} $, but $\omega
_{|m|, n} - \omega _{-|m|, n} \ll \omega _{|m|, n}$ \cite{IvZasAll}.
The same classification is valid for vortices for local EP FM
\cite{ISchMW}. Wysin had demonstrated  the direct correspondence of
gyroscopic character of vortex dynamics and doublet splinting
\cite{GaryMass}.

For an AFM vortex state particle each of two magnon branches, QFM
and QAF, produce a set of discrete modes with given $n$ and $m$,
however their properties are different compared to that for a FM
dot. The aforementioned formal Lorentz-invariance of spin dynamics
of AFM manifests itself for motion of a AFM vortex core: the
dynamical equation for the core coordinate $\vec X$ possess an
inertial term, $M_{\mathrm{v}}d^2\vec X / dt^2$, where the effective
vortex mass $M_{\mathrm{v}} = E_{\mathrm{v}} / c^2$
\cite{IvShekaPRL04,GaryMass}.  For this reason, the vortex core
dynamics is not a precession, as for the gyroscopic Thiele equation
for FM vortices \cite{Thiele,Huber,Sonin}, but rectilinear
oscillations, $\vec X(t) = \vec a\cos (\omega _{\mathrm{v}} t + \phi
_0 )$ degenerated with respect to the direction $\vec a$ and $\phi
_0 $, with the frequency $\omega _{\mathrm{v}} = \sqrt {\kappa /
M_{\mathrm{v}}} $, $\kappa $ determine the restoring force $\vec F =
- \kappa \vec X$ for the vortex. For EP AFM model with $W_m = 0$
such dynamics has been observed by direct numerical simulations
\cite{AFMvort91,AFMvort96}. For the vortex state particle with $R
> R_{\mathrm{crit}} $ the value of $\kappa $ is determined by the
demagnetizing field, $\kappa = 10\cdot 4 \pi
M^2_{\mathrm{DM}}L^2/9R$ \cite{GusIv}, and
\begin{equation} \label{v}
\omega _{\mathrm{v}} = \frac{2 cM_{\mathrm{DM}} \sqrt {10 L }}{3
\sqrt {AR\ln (\rho R / l_0 )} }\,.
\end{equation}

A simple estimate gives that $\omega _{\mathrm{v}} $, as for FM
vortex, is in subGHz region, but with different (approximately
square root, instead of linear for FM vortex) dependence  on the
aspect ratio $L / R$. The other modes from this set far from the
vortex core are characterized approximately by in-plane oscillations
of $\vec l$ and $\vec M$. As their frequencies are small, $\omega
\ll \gamma H_{\mathrm{DM}} $, for these modes the magnetization
$\vec {M}$ is determined mainly by the in-plane static contribution
\eqref{eq5}, and the formulae for the demagnetization field energy
for FM vortices can be used. For these modes the frequencies are of
the order of a few GHz, with approximately square root dependence on
the aspect ratio $L / R$, the details will be present elsewhere. The
absence of gyroscopical properties for the $\sigma$-ME is also
manifested in the fact that for a AFM vortex the modes with the
azimuthal numbers $m = \vert m\vert $ è $m = - \vert m\vert $ are
degenerated, i.e. splitting of doublets with $m = \pm \vert m\vert
$, typical for the FM vortex, is absent \cite{IvKolW,AFMvort96}.

For a vortex state AFM particle the high-frequency QAF branch of
magnons begets a set of discrete modes with frequencies of the order
of $\omega _\mathrm{g} $, i.e. hundreds GHz. For these modes far
from the vortex core oscillations of the vector $\vec {l}$ are out
of plane. For their description the dipole interaction is not
essential and the results obtained earlier for the vortex in EP AFM
\cite{IvKolW,AFMvort96} can be used. The mode frequencies $ \omega
_{n,m}$ are close to $ \omega _\mathrm{g}$, and the difference
$\omega _{n,m} -\omega _\mathrm{g} $ decreasing as the dot radius
increase as $c^2 /\omega _\mathrm{g} R^2 $, with one exception:
within the set of radially-symmetrical modes with $m = 0$ a
\emph{truly local} mode is present, with the localization area of
the order of $5l_0 $ and with the frequency $\omega _\mathrm{l} \sim
0.95\omega _\mathrm{g} $ independent on $R$ \cite{IvKolW}.

The usage of QAF modes for vortex state AFM particles, particularly
the truly local mode, would allow application of the idea of
magnonincs for higher frequencies till 0.3 THz. A developed theory
would be applied  for other AFM systems like a FM bilayer dot
containing two thin FM films with AFM interaction between them,
described by the field $H_{\mathrm{ex}}$. If $H_{\mathrm{ex}}$ is
large enough, $H_{\mathrm{ex}} > 4 \pi M_s$, anti-phase oscillations
of magnetic moments of the layers produce high frequency modes with
frequencies of order of $\sqrt{\gamma H_{\mathrm{ex}}
\omega_{m,n}^{FM}}$, where $\omega_{m,n}^{FM}$ are the frequencies
of modes for a single layer dot.

To conclude, for submicron particles of typical canted AFM the
ground state comprises topologically non-trivial spin distribution.
The magnetization of each sublattices $\vec {M}_1 $ and $\vec {M}_2
$ are characterized by a vortex state with a standard out of plane
structure, but the net magnetization $\vec {M}=\vec {M}_1 +\vec
{M}_2 $ form the planar vortex, where the magnetization in the
vortex center turns to zero. Vortex state AFM particles possess a
rich variety of normal magnon modes, from rectilinear oscillations
of the vortex core position with sub-GHz frequency till out of plane
modes with frequencies of order of hundreds GHz, including truly
local mode.

This work was supported by the grant \#219-09 from Ukrainian Academy
of Science.

\end{document}